\documentclass[reprint,
 superscriptaddress,
nofootinbib,
 aps,
pra,
twocolumn]{revtex4-2}

\usepackage[utf8]{inputenc} 
\usepackage[T1]{fontenc}    

\usepackage[english]{babel} 
\usepackage{lmodern}

\usepackage{hyperref}       
\usepackage{url}            
\usepackage{booktabs}       
\usepackage{amsfonts}       
\usepackage{nicefrac}       
\usepackage{microtype}      
\usepackage{graphicx}
\usepackage{xcolor, etoolbox}
\usepackage{amsmath,amsfonts,amsthm} 
\usepackage{mathtools}
\usepackage{ragged2e}
\usepackage{braket}      
\graphicspath{{./figures/}} 
\usepackage{enumitem}
\usepackage{xspace}
\usepackage{dsfont}
\usepackage{bm}
\usepackage{bbm}
\usepackage{dcolumn}
\usepackage{mathrsfs}
\usepackage{comment}

\usepackage{algorithmicx}
\usepackage{algpseudocode}
\usepackage{algcompatible}
\usepackage[linesnumbered,boxruled]{algorithm2e}
\newtheorem{theorem}{Theorem}
\newtheorem*{corollary}{Corollary}

\newtheorem*{informaltheorem}{Informal theorem}

\newcommand{\tr}{\text{Tr}}

\newcommand{\bs}[1]{\boldsymbol{#1}}
\newcommand{\mc}[1]{\mathcal{#1}}
\newcommand{\mcbs}[1]{\bs{\mc{#1}}}
\newcommand{\ketbra}[2]{\ket{#1}\bra{#2}}
\newcommand{\proj}[1]{\ketbra{#1}{#1}}

\hypersetup{
    colorlinks,
    linkcolor=blue,
    citecolor=blue,
    urlcolor=blue,
    breaklinks=true 
}

\newcommand{\norm}[1]{\left\|#1\right\|}

\begin{document}

\title{New random compiler for Hamiltonians via Markov Chains}

\author{Benoît Dubus}
    \email{Benoit.Dubus@ulb.be}
    \affiliation{Centre for Quantum Information and Communication, École polytechnique de Bruxelles, CP 165, Université libre de Bruxelles, 1050 Brussels, Belgium}

\author{Joseph Cunningham}
    \email{Joseph.Cunningham@ulb.be}
    \affiliation{Centre for Quantum Information and Communication, École polytechnique de Bruxelles, CP 165, Université libre de Bruxelles, 1050 Brussels, Belgium}

\author{Jérémie Roland}
    \email{Jeremie.Roland@ulb.be}
    \affiliation{Centre for Quantum Information and Communication, École polytechnique de Bruxelles, CP 165, Université libre de Bruxelles, 1050 Brussels, Belgium}

\begin{abstract}
Hamiltonian simulation is an important task in many quantum algorithms, for example in frameworks such as adiabatic quantum computation (AQC) or singular value transformation. In addition, the simulation of quantum systems is an important objective on its own. In many cases, the Hamiltonian can be decomposed as a linear combination of simpler terms; for instance, a sum of one- or two-qubit Hamiltonians for Ising models or a sum of time-independent Hamiltonians with time-dependent coefficients in AQC.
In this paper we develop a random compiler, similar to the first order randomized Trotter, or qDRIFT~\cite{campbellRandomCompilerFast2019}, leading to a new framework that is at the same time quite simple to implement and analyze and rather versatile as it supports a large class of randomization schemes as well as time-dependent weights. We first present the model and derive its governing equations. We then define and analyze the simulation error for a sum of two Hamiltonians, and generalize it to a sum of $Q$ Hamiltonians. We prove that the number of gates necessary to simulate the weighted sum of $Q$ Hamiltonians of norm $C$ during a time $T$ with an error less than $\epsilon_0$ grows as $\tilde{\mathcal{O}}\left(C^2T^2\epsilon_0^{-1}\right)$.
\end{abstract}

\maketitle

\let\clearpage\relax
\section{Introduction}
In this paper we consider new methods to simulate the evolution under a weighted sum of Hamiltonians for a time $T$. This problem falls in the scope of the simulation of the dynamics of quantum systems, one of the main applications of quantum computers as already noted by Feynman in 1972~\cite{feynmanSimulatingPhysicsComputers1982} and this has been proven relevant as Hamiltonian simulation in itself arises in many applications~\cite{lloydUniversalQuantumSimulators1996,sommaSimulatingPhysicalPhenomena2002,aspuru-guzikSimulatedQuantumComputation2005,lucasIsingFormulationsMany2014,childsQuantumAlgorithmSystems2017,robertResourceefficientQuantumAlgorithm2021,anLinearCombinationHamiltonian2023}.
An extensive review of the known methods for Hamiltonian simulation can be found in~\cite{zlokapaHamiltonianSimulationLowenergy2024}.

Assume that we need to simulate a target Hamiltonian $\bs{H}$ that can be decomposed as a convex combination of $Q$ Hamiltonians $\bs{H}_i$:
\begin{equation}
    \bs{H}=\sum_{i=1}^Q w_i \bs{H}_i,\label{Def:Ham1}
\end{equation}
where $\sum_i w_i=1,~w_i\geq0$. We assume that while $\bs{H}$ is hard to simulate directly, each elementary Hamiltonian is easy to implement. 

This problem arises in many instances such as for the simulation of $k$-local Hamiltonians, that can be decomposed as a sum of a large number of Hamiltonians acting on $k$ qubits, or in adiabatic quantum computing, that typically requires the implementation of a linear interpolation between an initial and a final Hamiltonian, and can therefore be decomposed as a linear combination of two time-independent Hamiltonians with time-dependent coefficients.

Usual approaches for such a problem are based on the Trotter-Suzuki product formula~\cite{suzukiFractalDecompositionExponential1990,suzukiFractalPathIntegrals1992,berryEfficientQuantumAlgorithms2007}, which in their simplest form lead to deterministic compilation schemes, but introducing randomness has been shown to be advantageous, including in the context of Hamiltonian simulation. Poulin \textit{et al.} exhibited the advantage of randomized schemes for the simulation of time-dependent Hamiltonians~\cite{poulinQuantumSimulationTimeDependent2011} then further literature showed that stochastic noise is less detrimental than coherent noise~\cite{wallmanNoiseTailoringScalable2016,kneeOptimalTrotterizationUniversal2015}. Successively, Campbell and Hastings proved that in Trotter-based models, randomness keeps errors smaller than in deterministic schemes~\cite{campbellShorterGateSequences2017,hastingsWeightReductionQuantum2017}. Finally, random Trotter schemes~\cite{childsFasterQuantumSimulation2019} and qDRIFT~\cite{campbellRandomCompilerFast2019} were introduced bringing new advancement in the performance.

In this paper, we introduce a scheme that alternates randomly between short evolutions under each Hamiltonian $\bs{H}_i$. The duration of each short evolution, and the next chosen Hamiltonian $\bs{H}_j$, are governed by Poisson processes of variable transition rates $A_{ij}$. Such a set of Poisson processes describes a continuous-time Markov chain. The parameters $A_{ij}$ define the randomization scheme and are the tunable parameters of this method. This method is similar to qDRIFT as both are randomization schemes based on Markov chains, the main difference being that qDRIFT works with discrete-time Markov chains whereas this new scheme considers continuous-time Markov chains. This allows us to describe the process using continuous differential equations rather than discrete difference equations. The complexity of the algorithms deduced from both methods are similar, but this method has the advantage that it allows randomization of the time steps, which could be useful for the implementation of adiabatic algorithms, since such algorithms typically require a carefully chosen schedule to run efficiently~\cite{rolandQuantumSearchLocal2002a,cunninghamEigenpathTraversalPoissonDistributed2024}.

This paper keeps the same scope as~\cite{campbellRandomCompilerFast2019}, as our techniques falls under Trotter-based methods that do not require additional ancilla qubits. We will therefore not compare our techniques to methods such as LCU~\cite{chakrabortyImplementingAnyLinear2024} or qSWIFT~\cite{nakajiHighOrderRandomizedCompiler2024}, using ancilla qubits, as those have been proven to lead to further improvements than simple Trotter-based methods \cite{berryBlackboxHamiltonianSimulation2012,berryHamiltonianSimulationNearly2015,berrySimulatingHamiltonianDynamics2015,babbushEncodingElectronicSpectra2018,lowHamiltonianSimulationQubitization2019}. Moreover, we will also only consider analytical results and therefore will not compare our methods to approaches based on numerical results that are often more optimistic~\cite{babbushChemicalBasisTrotterSuzuki2015,poulinTrotterStepSize2015,childsFirstQuantumSimulation2018}.

In Section~\ref{Sec:Description}, we give a detailed description of the model and derive the differential equations governing its evolution. In Section~\ref{Sec:General Results} we prove that this system approximately simulates the evolution under the target Hamiltonian~\eqref{Def:Ham1} for any input state. In Section~\ref{Sec:Error}, we derive an upper-bound on the error made during the simulation depending on the tunable coefficients $A_{ij}$ and we use them in Section~\ref{Sec:Complexity} to derive upper-bounds on the number of gates necessary to implement such Hamiltonian simulation in a gate-based model. In particular we derive two theorems and their corollaries:
\begin{informaltheorem}
Assuming that each individual Hamiltonian can be applied for a finite time with $O(1)$ gates, the evolution under $\bs{H}=\sum_{i=1}^Q \bs{H}_i$ for a duration $T$ can be simulated with an error smaller than $\epsilon_0$ using a number of gates scaling as
\begin{equation}
		\bar{G}=\tilde{O}\left(\frac{c^2 T^2}{\epsilon_0}\right),
	\end{equation}
where $c=\sum_i \left\|\bs{H}_i\right\|_\infty$.
\end{informaltheorem}

\section{Description of the model}\label{Sec:Description}
Random Trotterization consists in simulating the Hamiltonian 
\begin{equation}
    \bs{H}:=\sum_{i=1}^Q w_i \bs{H}_i
\end{equation}
by alternating randomly between the Hamiltonians $ \bs{H}_i$. Using the Trotter-Suzuki formula, one can show that this process approximates evolution under $H$. However the calculations are complex. In this paper, we give more attention to the random process governing the alternation; specifically we will consider processes based on Markov chains.

The general procedure is described by Algorithm~\ref{procedure}. The principle is to sample a realization of a continuous-time Markov chain on a graph with $Q$ nodes (corresponding to the $Q$ Hamiltonians $H_i$), and translate that realization as a quantum circuit. Then, the transition rates defining the Markov chain $A_{ij}(t)$ are computed thanks to the parameter $\lambda$, chosen large enough to ensure small errors, as discussed in Section~\ref{Sec:Complexity}, and the desired instantaneous probability distribution $\{w_i(t)\}$. Said Markov chain, is prepared in a way that for all time the probability to be in node $i$ is $w_i$. A realization of that Markov chain is sampled and throughout the evolution, the Hamiltonian applied on the quantum system, is the one corresponding to the current state of the chain (when the chain is in node $i$, the $i^{th}$ Hamiltonian is applied unto the quantum system).


\begin{algorithm}\caption{General procedure}\label{procedure} 

\SetKwInOut{Input}{Input}

\Input{A total duration \(T\), \\ A set of Hamiltonians \(\{\bs{H}_i\}\), \\ A set of weights \(\{w_i(t)\}\), \\ A total transition rate $\lambda$.}
$C\leftarrow \max_i\|\bs{H}_i\|_\infty$ \;

$A_{ij}\leftarrow \dot{w}_j+\lambda w_j$\;
$T_{list}\leftarrow$ Sample the list of tuples of nodes and time intervals of a realization of a Continuous Time Markov Chain with transition rates $A_{ij}$ and a duration $T$\;

$V_{list}\leftarrow\{\}$ (Empty list)\;

\For{$(k,\tau)$ in $T_{list}$}{
	Append $e^{-iH_k\tau}$ to $V_{list}$\;
} 

\textbf{Return: }$V_{list}$\;
\end{algorithm}

\subsection{Quantum Description of Classical Markov chains}

A classical continuous-time Markov chain is a stochastic process on a graph of $Q$ accessible nodes, with oriented weighted edges. The state of this system at time $t$ is described by a probability vector $\vec{p}$ (of dimension $Q$), where $p_i$ denotes the probability to be in node $i$.  The matrix $A$ of size $Q\times Q$, where $A_{ij}$ is the weight of the edge from $i$ to $j$, represents the transition rates, \textit{i.e.} the probability of transition from node $i$ to $j$ for an infinitesimal time step $dt$ is $A_{ij} dt$. The coefficient $A_{ij}$ is positive for all $i\neq j$, and $A_{ii}=-\sum_{j\neq i}A_{ij}$. In summary, the evolution of the system is governed by Equation~\cite{frigessiMarkovChains2011,yinContinuousTimeMarkovChains2013}:
\begin{equation}
    \partial_t \vec{p}=A^T\vec{p}.\label{Eq:Des_Evol_MC_Class}
\end{equation}
It conserves the total probability as expected. For a time-dependent continuous-time Markov chain, the matrix $A$ will be itself time-dependent, but we will impose that it is differentiable.

The matrix $A^T$ has several properties due to its structure~\cite{frigessiMarkovChains2011,yinContinuousTimeMarkovChains2013}. It is diagonalizable, all its eigenvalues have non-positive real parts, if there are complex eigenvalues then they come in pairs of complex conjugates, and there is at least one null eigenvalue. In this paper we will assume that the null eigenvalue is not degenerate. In that case, we will call the eigenvector associated with the eigenvalue $0$ the stationary vector $\vec{q}$.

In order to make a quantum description of this classical process, we will encode the probability vector $\vec{p}$ as a density matrix $\boldsymbol{P}:=\sum_i p_i \proj{i}$, that is, a quantum state in the Hilbert space $\mathcal{H}_{\mathrm{Mark}}$ isomorphic to $\mathbb{C}^Q$, where $\ket{i}$ represents the node $i$ of the Markov chain. The classicality of this system lies in the preservation of the diagonality with regard to the basis $\{\ket{i}\}$ along the evolution.

The evolution of the system can now be represented by the trace-preserving equation
\begin{equation}
\begin{aligned}
    \partial_t \boldsymbol{P}&=\sum_{i,j=1}^Q A_{ij} \ketbra{j}{i}\boldsymbol{P}\ketbra{i}{j},\label{Eq:Des_Evol_MC_Quant}\\
    &=\sum_{i\neq j}^Q A_{ij} \braket{i|\boldsymbol{P}|i}\left(\ketbra{j}{j}-\ketbra{i}{i}\right),
    \end{aligned}
\end{equation}
or equivalently, defining the jump operators $\boldsymbol{L}_{i\rightarrow j}:=\ketbra{j}{i}$, the Lindbladian equation
\begin{equation}
    \partial_t \boldsymbol{P}=\sum_{i\neq j}^QA_{ij}\left(\boldsymbol{L}_{i\rightarrow j}\boldsymbol{P} \boldsymbol{L}_{i\rightarrow j}^\dagger -\frac{1}{2} \left\{\boldsymbol{L}_{i\rightarrow j}^\dagger \boldsymbol{L}_{i\rightarrow j}, \boldsymbol{P}\right\} \right).\label{Eq:Des_Evol_MC_Lind}
\end{equation}

As mentioned before, for \textit{classical} states, \textit{i.e.}, when $\boldsymbol{P}$ is diagonal in the basis $\{\ket{i}\}$, equations~\eqref{Eq:Des_Evol_MC_Class},~\eqref{Eq:Des_Evol_MC_Quant} and~\eqref{Eq:Des_Evol_MC_Lind} are equivalent. 

\subsection{Control of the quantum system}
We now consider a quantum system, typically $n$ qubits, lying in the Hilbert space $\mathcal{H}_{\mathrm{Quant}}$  isomorphic to $\mathbb{C}^{2^n}$. As the complete system is composed of this quantum system and the Markov chain, it lies in the Hilbert space $\mathcal{H}_{\mathrm{Tot}}=\mathcal{H}_\mathrm{Quant}\otimes\mathcal{H}_{\mathrm{Mark}}$.

The quantum system is controlled by the Markov chain, \textit{i.e.}, if the Markov chain is in node $\ket{i}$, then the Hamiltonian $\bs{H_i}$ is applied onto the quantum system. If we consider the density matrix of the coupled system as $\mcbs{P}\in\mathrm{End}\left(\mathcal{H}_{\mathrm{Tot}}\right)$, then this controlled Hamiltonian can be expressed as a Hamiltonian $\bs{\mathcal{H}}:=\sum_{i=1}^Q\bs{H_i} \otimes \proj{i}$. 

The complete governing Lindbladian equation for the whole system is
\begin{equation}\label{Eq:Des_evol_tot_sys}
\begin{aligned}
    \partial_t\mcbs{P}=&-i\left[\mcbs{H},\mcbs{P}\right]\\&+\sum_{i\neq j}^QA_{ij}\left(\mcbs{L}_{i\rightarrow j}\mcbs{P} \mcbs{L}_{i\rightarrow j}^\dagger -\frac{1}{2} \left\{\mcbs{L}_{i\rightarrow j}^\dagger \mcbs{L}_{i\rightarrow j}, \mcbs{P}\right\} \right)\\
    =&-i\left[\mcbs{H},\mcbs{P}\right]+\sum_{i, j=1}^QA_{ij}\braket{i|\mcbs{P}|i}\left(\proj{j}-\proj{i}\right),
    \end{aligned}
\end{equation}
where the jump operators are $\bs{\mathcal{L}}_{i\rightarrow j}:=\bs{\mathds{1}}\otimes \bs{L}_{i\rightarrow j}$, still representing the jumps in the state of the Markov chain.
With the first expression, we can prove that if $\mcbs{P}$ is block-diagonal, then it is left block-diagonal by this evolution (preserving the classicality of the Markov chain), and this leads to the second expression.

Tracing out the quantum system, we recover Equation~\eqref{Eq:Des_Evol_MC_Lind}, \textit{i.e.} the evolution of a classical Markov chain. 

In practice, there is no need to physically implement the Markov chain, one can first simulate it in order to find one of its realizations, then use it to define the sequence of Hamiltonians that will be applied to the quantum system.  This description of the process will therefore only be used to prove analytical bounds on simulation errors.


\subsection{Interpretation} 
The classical Markov chain evolution equation~\eqref{Eq:Des_Evol_MC_Class} can be rewritten as
\begin{equation}
    \partial_t p_i=\sum_{j\neq i}^QA_{ji} p_j-\sum_{j\neq i}^QA_{ij}p_i\qquad\forall i,\label{eq:MCInterp}
\end{equation}
therefore the evolution of the probability of presence in the node $i$ is a sum of incoming fluxes from all other nodes and outgoing fluxes. 

We can find a similar equation for the quantity $\bs{\rho}_i:=\braket{i|\mcbs{P}|i}$. This density matrix is not normalized to one as $\tr\left(\bs{\rho}_i\right)=p_i$. The density matrix $\bs{\rho}_i$ therefore conveys the probability of the Markov chain to be in the node $i$, and renormalising it as $\bs{\rho}_i/p_i$ describes the average state of the quantum system over all realizations of the Markov chain, given that the Markov chain is in node $i$ at time $t$. Equation~\eqref{Eq:Des_evol_tot_sys} can be rewritten as
\begin{equation}
    \partial_t \bs{\rho}_i=-i\left[\bs{H}_i,\bs{\rho}_i\right]+\sum_{j\neq i}^QA_{ji} \bs{\rho}_j-\sum_{j\neq i}^QA_{ij}\bs{\rho}_i\qquad\forall i.
\end{equation}
We see that the marginal density matrix in the node $i$ evolves under the Hamiltonian $\bs{H}_i$ with incoming and outgoing density fluxes.
\subsection{Balanced Scheme}
In this paper we will focus on a specific randomization scheme that we will refer to as balanced scheme, and therefore a particular Markov chain, the initial-node independent chain. In this case the transition rate from $i$ to $j$ does not depend on $i$, therefore $\forall i~:~A_{ij}=a_j$ where $a_j$ is then the transition rate from any node to node $j$.

From Equation~\eqref{eq:MCInterp}, we see that the evolution of the Markov chain is governed by
\begin{equation}
    \partial_t p_i=\sum_{j=1}^Qa_i p_j-\sum_{j=1}^Qa_jp_i.
\end{equation}
Introducing the total transition rate as $\lambda:=\sum_{i=1}^Q a_i$, the evolution of the Markov chain is now
\begin{equation}
    \Rightarrow \partial_t \vec{p}=\vec{a}-\lambda\vec{p}.
\end{equation}
We can decompose the probability vector in the stationary probability vector $\vec{q}=\vec{a}/\lambda$ and a bias vector $\vec{r}=\vec{p}-\vec{q}$.

This equation can explicitly be solved as 
\begin{equation}
    \vec{p}(t)=\vec{q}(t)+\vec{r}(0)e^{-\int_0^t\lambda d\tau}-\int_0^t\partial_t q(\tau)e^{-\int_\tau^t\lambda d\tau'}d\tau.
\end{equation}

For this particular scheme, Equation~\eqref{Eq:Des_evol_tot_sys} for the evolution of the total system becomes
\begin{equation}
    \partial_t\mcbs{P}=-i\left[\mcbs{H},\mcbs{P}\right]-\lambda\left(\mcbs{P}-\tr_M\left(\mcbs{P}\right)\otimes\sum_i q_i\proj{i}\right),
\end{equation}
Those equations are rather simpler and allow us to easily prove bounds on simulation errors and gate costs.

\section{General results}\label{Sec:General Results}
In general we will focus on the simulation of a Hamiltonian $\bs{H}=\sum_{i=1}^Q w_i\bs{H}_i$, where $w_i\geq 0$ and $\sum_iw_i=1$ such that $\vec{w}$ can be interpreted as a probability vector. We define the exact evolution channel $\bs{E}_t$ that sends an initial density matrix $\sigma$ onto the one obtained after evolution during time $t$ under the Hamiltonian $\bs{H}=\sum_{i=1}^Q w_i\bs{H}_i$, \textit{i.e.}:
\begin{equation}
    E_t:\left(\sigma,t\right)\rightarrow \mathcal{T}\left(e^{-i\int_0^t\bs{H}d\tau}\right)\sigma\mathcal{T}\left(e^{i\int_0^t\bs{H}d\tau}\right).
\end{equation}
Similarly, we define the simulation channel $\bs{S}_t$ implemented by the procedure described in section~\ref{Sec:Description} for a time $t$. In particular, the probability to be in each node of the Markov chain at each time is $\vec{p}(t)=\vec{w}(t)$, this is satisfied by setting $A_{ij}=\partial_t w_j+lambda w_j$. Since $A_{ij}$ must be nonnegative, this choice is valid for large enough lambda as soon as $-\frac{\partial_t w_j}{w_j}$ is upper bounded by a constant. This assumption is trivially satisfied for time-independent or smoothly varying $w_i$ as long as they do not go to $0$ during the evolution\footnote{A typical case where this would not be satisfied is at the end of an adiabatic evolution. In such cases, it is also possible to consider Markov chains with more complex coefficients $A_{ij}$. If there is only two nodes we can also consider unbounded $\lambda$, and a more precise analysis allows us to demonstrate that the mean number of gates stays finite.}.

Formally, these channels map an initial density matrix $\sigma$ to a density matrix with an additional register for the Markov chain, that is, $\mcbs{P}=\sigma\otimes \sum_i w_i(0)\proj{i}$, let this state evolve for a time $t$, then trace out the Markov chain.

Tracing the Markov chain out of Equation~\eqref{Eq:Des_evol_tot_sys}, the average state of the quantum system over all realizations of the Markov chain, that is, $\bs{\rho}:=\tr_M\left(\mcbs{P}\right)$, evolves as
\begin{equation}\label{Eq:Evol_rho}
    \partial_t \bs{\rho}=-i\left[\bs{H},\bs{\rho}\right]-i\tr_M\left(\left[\mcbs{H},\mcbs{R}\right]\right), 
\end{equation}
 where $\mcbs{R}:=\mcbs{P}-\bs{\rho}\otimes \bs{W}(t)$ is a difference between density matrices, the density matrix $\mcbs{P}$ of the whole system and its product state approximation, and is thus Hermitian but not semidefinite positive, and $\bs{W}(t):=\sum_iw_i(t)\proj{i}$. We can further show that $\forall i~:~\tr\left(\braket{i|\mcbs{R}|i}\right)=0$. 
Since the channel $\bs{E}_T$ yields evolution under Hamiltonian $\bs{\rho}$, Equation~\eqref{Eq:Evol_rho} immediately implies that this channel is well approximated by $\bs{S}_T$ if $\|\mcbs{R}\|$ is small (for some norm). Inserting the evolution of $\mcbs{R}$ and $\bs{\rho}$ in~\eqref{Eq:Des_evol_tot_sys}, we find
\begin{equation}
\begin{aligned}
    \partial_t\mcbs{R}=-&i\left(\left[\mcbs{H},\mcbs{R}\right]-\tr_M\left(\left[\mcbs{H},\mcbs{R}\right]\right)\otimes\bs{W}\right)\\
    &+\sum_{i\neq j}A_{ij}\left(\mcbs{L}_{i\rightarrow j}\mcbs{R} \mcbs{L}_{i\rightarrow j}^\dagger -\frac{1}{2} \left\{\mcbs{L}_{i\rightarrow j}^\dagger \mcbs{L}_{i\rightarrow j}, \mcbs{R}\right\} \right)\\
    &-i\left[\mcbs{H}-\bs{H}\otimes\bs{\mathds{1}},\bs{\rho}\otimes \bs{W}\right],\label{Eq:Evol_STP_Gen}
\end{aligned}
\end{equation}

The first term in this equation shows that $\mcbs{R}$ evolves under the Hamiltonian $\mcbs{H}$. However, $\mcbs{R}$ is supposed to have a partial trace (tracing out the Markov chain) null, and this is ensured by the second term. More precisely, we can define the projector
\begin{equation}
    \bs{\Pi}_{\vec{w}}:\mcbs{P}\rightarrow\mcbs{P}-\tr_M\left(\mcbs{P}\right)\otimes\bs{W},
\end{equation}
which is such that $\tr_M\left(\bs{\Pi}_{\vec{w}}\mcbs{P}\right)=0$ for any $\mcbs{P}$,
and write the first two terms as
\begin{equation}
-i\bs{\Pi}_{\vec{w}}\left(\left[\mcbs{H},\mcbs{R}\right]\right)
\end{equation}
The second line of~\eqref{Eq:Evol_STP_Gen} represents the exchange between components $\braket{i|\mcbs{R}|i}$ of $\mcbs{R}$ due to the jump processes of the Markov chain. The third line is a bias source, arising from the difference between the Hamiltonian $\bs{H}_i$ actually applied at each node, and the average Hamiltonian $\bs{H}$; thus the closer to each other the Hamiltonians will be, the smaller the error will be.

From Equation~\eqref{Eq:Evol_rho}, we see that the norm of the bias $\bs{\Delta\rho}=\left(\bs{E}_T-\bs{S}_T\right)\bs{\rho}$ can be bounded as
\begin{equation}
    \left\|\bs{\Delta\rho}\right\|\leq\int_0^t\left\|\tr_M\left(\left[\mcbs{H},\mcbs{R}\right]\right)\right\|,\label{Eq:Ev_NormDelta0}
\end{equation}
independently of the norm considered. In the following sections, we will bound the norm of the bias for balanced schemes, and prove that in some limits it vanishes, \textit{i.e.} the estimator is unbiased.

\section{Error for Balanced Schemes}\label{Sec:Error}
For balanced schemes, Equation~\eqref{Eq:Evol_STP_Gen} 
reduces to
\begin{equation}\label{Eq:Evol_ST_Sym}
    \partial_t\mcbs{R}=-i\bs{\Pi}_{\vec{w}}\left(\left[\mcbs{H},\mcbs{R}\right]\right)-\lambda\mcbs{R}-i\left[\mcbs{H}-\bs{H}\otimes\bs{\mathds{1}},\bs{\rho}\otimes \bs{W}\right],
\end{equation}
which will simplify the analysis.

In the following section, we will use certain conventions and properties extensively:
\begin{enumerate}
\item The initial state of the full system is the product state  $\mcbs{P}(0)=\bs{\rho}(0)\otimes\bs{W}(0)$. As $\bs{W}(t)$ is diagonal for all $t$, the initial state is block-diagonal.
\item The evolution of the Markov chain preserves the block-diagonality, as discussed in previous sections, so that $\mcbs{P}(t)$ is block-diagonal for all times t.
\item Using its definition, and the diagonality of $\bs{W}$,  $\mcbs{R}=\mcbs{P}-\tr_M(\mcbs{P})\otimes\bs{W}$ is also block-diagonal for all $t$, and $\mcbs{R}(0)=0$.
\item As tracing the quantum system out of equation~\eqref{Eq:Des_evol_tot_sys}, returns the equation~\eqref{Eq:Des_Evol_MC_Lind}, we have that $\tr_Q(\mcbs{P})=\bs{W}$.
\item Therefore we find the $\tr_M(\mcbs{R})=0$ and $\tr(\braket{i|\mcbs{R}|i})=0$.
\end{enumerate}

\subsection{The 2-nodes case}
First, let us note that any 2-node Markov chain corresponds to a balanced scheme. We can define $\bs{\tilde{\rho}}=\braket{1|\mcbs{R}|1}$, and using that $\tr_M(\mcbs{R})=0$, we can write $\mcbs{R}=\bs{\tilde{\rho}}\otimes\left(\proj{1}-\proj{2}\right)$. Equation~\eqref{Eq:Ev_NormDelta0} now becomes
\begin{equation}
\norm{\bs{\Delta\rho}}\leq\int_0^t\norm{\left[\bs{\Delta H},\bs{\tilde{\rho}}\right]}
\end{equation}
for any norm $\|\cdot\|$ with $\bs{\Delta H}:= \bs{H}_1-\bs{H}_2$ and Equation~\eqref{Eq:Evol_ST_Sym} gives
\begin{equation}
    \partial_t\bs{\tilde{\rho}}=-i\left[\bs{\tilde{H}},\bs{\tilde{\rho}}\right]-\lambda\bs{\tilde{\rho}}-iw_1w_2\left[\bs{\Delta H},\bs{\rho}\right],
\end{equation}
where $\bs{\tilde{H}}:= w_2\bs{H}_1+w_1\bs{H}_2$. This equation can be solved explicitly, and using that $\bs{\tilde{\rho}}(0)=\bs{0}$, the solution is 
\begin{equation}
\begin{aligned}
    \bs{\tilde{\rho}}=-i\int_0^t& w_1w_2e^{-\int_\tau^t\lambda d\tau'}\mathcal{T}\left(e^{-i\int_\tau^t \bs{\tilde{H}}d\tau'}\right)\\&\left[\bs{\Delta H},\bs{\rho}\right]\mathcal{T}\left(e^{i\int_\tau^t\bs{\tilde{H}}d\tau'}\right)d\tau.
    \end{aligned}
\end{equation}
This directly implies that
\begin{equation}
    \norm{\bs{\tilde{\rho}}}\leq\int_0^t w_1w_2e^{-\int_\tau^t\lambda d\tau'}\norm{\left[\bs{\Delta H},\bs{\rho}\right]}d\tau.
\end{equation}
Now, using the Schatten $p$-norms $\|\cdot\|_p$, assuming $\forall t,\forall p\leq1~:~\norm{\bs{\rho}}_p\leq 1$:
\begin{equation}
\begin{aligned}
    \norm{\bs{\Delta\rho}}_p\leq&4\int_0^t\norm{\Delta H}_\infty\\&\qquad\int_0^\tau e^{-\int_{\tau'}^\tau\lambda d\tau''}w_1w_2\norm{\bs{\Delta H}}_\infty d\tau'd\tau.
    \end{aligned}
\end{equation}
This is an upper-bound on the simulation error, as by definition $\norm{\mcbs{E}_T-\mcbs{S}_T}_{p\rightarrow p}:=\sup_{\norm{\bs{\rho}}_p\leq 1}\norm{\bs{\Delta\rho}(T)}_p$.

Assuming, for the sake of simplicity, that $\bs{H_1}, \bs{H_2}$ and $\lambda$ are constant and finite, we get
\begin{equation}
    \norm{\bs{E}_T-\bs{S}_T}_{p\rightarrow p}\leq\frac{4T}{\lambda}\sup_T(w_1w_2)\norm{\bs{\Delta H}}_\infty^2.\label{Eq:Norm2Node}
\end{equation}
In the limit of $\lambda\rightarrow \infty$ this value goes to 0, and the estimator is unbiased.

\subsection{The $n$-node balanced case}
Using similar calculations, detailed in Appendix~\ref{App:1}, this result can be extended to a Hamiltonian with $Q$ terms $\bs{H}=\sum_{i=1}^Q w_i \bs{H}_i$:
\begin{equation}
    \norm{\bs{E}_T-\bs{S}_T}_{p\rightarrow p}\leq \frac{4Q^{\frac{p-1}{p}}\norm{\vec{w}}_{p}\norm{\mcbs{H}-\bs{H}\otimes\bs{\mathds{1}}}_\infty\norm{\mcbs{H}}_\infty T}{\lambda-2Q^{\frac{p-1}{p}}\norm{\mcbs{H}}_\infty\norm{\vec{w}}_p}.
\end{equation}
This equation can be further simplified in special cases. Mainly, if $w_i=\frac{1}{Q}$ or if $p=1$ (as we will consider in Section~\ref{Sec:Complexity}), then the product $Q^{\frac{p-1}{p}}\norm{\vec{w}}_{p}$ simplifies to one. In that case, assuming that the norm of each Hamiltonian in the sum is upper-bounded by $C$, \textit{i.e.} $\norm{\bs{H}_i}_\infty\leq C$ for all $i$, the simulation error becomes
\begin{equation}
    \norm{\bs{E}_T-\bs{S}_T}_{p\rightarrow p}\leq \frac{8C^2 T}{\lambda-2C}.\label{Eq:NormNNode}
\end{equation}
This does not depend on $Q$, and is similar to the bound with 2 terms.

\section{Complexity}\label{Sec:Complexity}
In this section, we consider applying this protocol in a gate-based model, where a realization of the Markov process dictates a sequence of Hamiltonians applied for given times, or equivalently a sequence of unitaries, further implemented as a sequence of gates. In that case, Equations~\eqref{Eq:Norm2Node} and~\eqref{Eq:NormNNode} lead to bounds on the number of gates needed to simulate the target Hamiltonian with an error bounded by $\epsilon$. We will develop two cost-models, the first one considers that each unitary can be implemented exactly, the second-one that each unitary can only be applied approximately at a cost that is proportional to the logarithm of the approximation error.

In this section we assume $\lambda$ constant, finite and that the $w_i$ vary slowly as discussed in section~\ref{Sec:General Results}.
\subsection{Perfect gates}
Let us consider that the evolution under Hamiltonian $H_i$ for time $\tau$ can be performed exactly using $\alpha+\beta C\tau$ gates. On average, the Markov chain will spend $(\lambda- a_i)^{-1}$ time-steps in the $i^{\textrm{th}}$ node. Each time-step has an expected length of $w_iT$  and so the Markov chain will be in the node $i$ for a duration of $w_iT(\lambda -a_i)$, on average. From this, we conclude that the expected number of gates is:
\begin{equation}
    \bar{G}=\sum_i w_i\left(\alpha(\lambda-a_i)+\beta C\right)T\leq T(\alpha\lambda+\beta C).
\end{equation}

$\bs{E}_T$ and $\bs{S}_T$ should be close, as the second is an estimator of the first, and therefore we will impose\footnote{Where $F$ is the fidelity between the two channels, \textit{i.e.}, the minimum fidelity of the outputs of both channels for the same input} $F(\bs{E}_T,\bs{S}_T)\geq1- \epsilon_0$, which is ensured by $\norm{\bs{E}_T-\bs{S}_T}_{1\rightarrow 1}\leq2\epsilon_0$\footnote{We used Fuchs–van de Graaf~\cite{fuchsCryptographicDistinguishabilityMeasures1999} inequalities that states
\begin{equation}
1-F(\bs{\psi},\bs{\rho})\leq\frac{1}{2}\norm{\bs{\psi}-\bs{\rho}}_1,
\end{equation}
for $\bs{\psi}$ a pure state. We use that $\bs{E}_T$ is pure, as the exact process is an Hamiltonian evolution, and that the initial state is pure.}. This implies 
\begin{equation}
    \frac{8C^2 T}{\lambda-2C}\leq 2\epsilon_0\Rightarrow \lambda\geq \frac{4C^2 T}{\epsilon_0}+2C.
\end{equation}
and therefore we can conclude our first theorem:
\begin{theorem} For any $\bs{H}_i, ~w_i$ such that  $\norm{\bs{H}_i}_\infty\leq C$,  $\sum_{i=1}^Q w_i=1$, and $w_i\geq0$, for any $\epsilon_0>0$, if each elementary Hamiltonian $ \bs{H}_i$ can be applied during any time-step $\tau$ perfectly at gate-cost $\alpha + \beta C\tau$, then there exits a $\lambda$ such that Algorithm~\ref{procedure} simulates $\bs{H}=\sum_{i=1}^Q w_i \bs{H}_i$ for a duration $T$ with an error smaller than $\epsilon_0$, with an expected number of gates
\begin{equation}
    \bar{G}\leq\frac{4\alpha C^2 T^2}{\epsilon_0}+(2\alpha+\beta) C T.
\end{equation}
\end{theorem}
The dominating term is the one independent of the length of the time-step. This is natural as each time-step should be short. 

Note that for this result, we are simulating $\bs{H}=\sum_{i=1}^Q w_i \bs{H}_i$ where the $w_i$ are normalized to one, thus hiding the dependency in $Q$. However, simulating $\sum_{i=1}^Q \bs{H}_i$ requires to renormalize the terms, that is, we simulate 
\begin{equation}
\bs{H}=\sum_{i=1}^Q w_i \bs{\tilde{H}}_i
\end{equation}
with $w_i:=\frac{\norm{\bs{H}_i}_\infty}{\sum_j\norm{\bs{H}_j}_\infty}$
and $\tilde{H}_i:=\frac{\sum_j\norm{\bs{H}_j}_\infty}{\norm{\bs{H}_i}_\infty}\bs{H}_i$,
such that $\norm{\tilde{H}_i}_\infty=\sum_j \norm{H_j}_\infty\forall i$.
\begin{corollary}
Assuming that each individual Hamiltonian can be applied for a finite time with $O(1)$ gates, the evolution under $\bs{H}=\sum_{i=1}^Q \bs{H}_i$ for a duration $T$ can be simulated with an error smaller than $\epsilon_0$ using a number of gates scaling as
\begin{equation}
		\bar{G}=\tilde{O}\left(\frac{c^2 T^2}{\epsilon_0}\right),
	\end{equation}
where $c=\sum_i \left\|\bs{H}_i\right\|_\infty$.
\end{corollary}

This allows us to compare our results with the state of the art. The following notations are used: $C:=\max_i \norm{H_i}_\infty$, and $c=\sum_i\norm{H_i}_\infty$. We find that the expected number of gates in our case has the exact same scaling as qDRIFT (see Table~\ref{tab:compStateArt}).

\begin{table}[h]
    \centering
    \caption{Comparison with the state of the art, this table is adapted from~\cite{campbellRandomCompilerFast2019}.}
    \begin{tabular}{|l|c|}
        \hline
        \textbf{Protocol} & \textbf{Gate count} (upper bound) \\ 
        \hline
        $1^{st}$ order Trotter DET & $O\left(Q^3(C T)^2\epsilon_0^{-1}\right)$ \\
        $2^{nd}$ order Trotter DET & $O\left(Q^{\frac{5}{2}}(C T)^{\frac{3}{2}}\epsilon_0^{-\frac{1}{2}}\right)$ \\
        $(2k)^{th}$ order Trotter DET & $O\left(Q^{\frac{4k+1}{2k}}(C T)^{\frac{2k+1}{2k}}\epsilon_0^{-\frac{1}{2k}}\right)$ \\
        \hline
        $(2k)^{th}$ order Trotter RANDOM & $O\left(Q^2(C T)^{\frac{2k+1}{2k}}\epsilon_0^{-\frac{1}{2k}}\right)$ \\
        \hline
        qDRIFT (general result) & $O\left((c T)^2\epsilon_0^{-1}\right)$ \\
        qDRIFT (when $c = QC$) & $O\left(Q^2(C T)^2\epsilon_0^{-1}\right)$ \\
        qDRIFT (when $c=\sqrt{Q}C$) & $O\left(Q(C T)^2\epsilon_0^{-1}\right)$ \\
        \hline
        Our (general result) & $O\left((c T)^2\epsilon_0^{-1}\right)$ \\
        Our (when $c = QC$) & $O\left(Q^2(C T)^2\epsilon_0^{-1}\right)$ \\
        Our (when $c=\sqrt{Q}C$) & $O\left(Q(C T)^2\epsilon_0^{-1}\right)$ \\
        \hline
    \end{tabular}
    \label{tab:compStateArt} 
\end{table}

\subsection{Imperfect gates}
Let us now use a more realistic model, where we can apply each $H_i$ for a time-step $\tau$, and for any error $\epsilon_1$ at a cost $(\alpha+\beta C\tau)\ln{\epsilon_1^{-1}}$. In that case, the expected number of gates is 
\begin{equation}
    \bar{G}\leq(T(\alpha\lambda+\beta C)\ln{\epsilon_1^{-1}}).\label{Eq:gate_count_imperfect2}
\end{equation}
and the total error made during the simulation will have an additional term $\lambda T \epsilon_1$, as each stay on a node induces an error $\epsilon_1$, so that
\begin{equation}
    \norm{\bs{E}_T-\bs{S}_T}_{p\rightarrow p}\leq \frac{8C^2 T}{\lambda-2C}+\lambda T \epsilon_1. 
\end{equation}
Once again, using Fuchs-van de Graaf inequalities, a sufficient condition to have final fidelity $\epsilon_0$ is to impose that
%
\begin{equation}
\begin{aligned}
&&     \frac{8C^2 T}{\lambda-2C}+\lambda T \epsilon_1&\leq 2\epsilon_0\\
&\Leftrightarrow &
\epsilon_1&\leq\frac{1}{\lambda T}\left(2\epsilon_0-\frac{8C^2T}{\lambda-2C}\right)
     \end{aligned}
\end{equation}

Saturating this equation, and setting
\begin{equation}
\lambda=\frac{4C^2T}{\epsilon_0}\left(1+\left(\ln\frac{8C^2T^2}{\epsilon_0}\right)^{-1}\right)+2C,
\end{equation}
in Equation~\eqref{Eq:gate_count_imperfect2} then leads to our second theorem.
\begin{theorem}
For any $\bs{H}_i, ~w_i$ such that  $\norm{\bs{H}_i}_\infty\leq C$,  $\sum_{i=1}^Q w_i=1$, and $w_i\geq0$, for any $\epsilon_0>0$, if each elementary Hamiltonian $ \bs{H}_i$ can be applied during any time-step $\tau$ with an error smaller than $\epsilon_1$ at gate-cost $\left((\alpha + \beta C\tau)\ln\left(\epsilon_1^{-1}\right)\right)$, then there exits a $\lambda$ such that Algorithm~\ref{procedure} simulates $\bs{H}=\sum_{i=1}^Q w_i \bs{H}_i$ for a duration $T$ with an error smaller than $\epsilon_0$, with an expected number of gates:
\begin{equation}
 \bar{G}=O\left(CT\left(\frac{\alpha C T}{\epsilon_0}+\beta\right)\ln\left(\frac{CT}{\epsilon_0}\right)\right).
\end{equation}
\end{theorem}

\section{Conclusion and Outlook}
In this paper, we introduced a new approach for simulating the evolution of quantum systems governed by a weighted sum of Hamiltonians using a new randomized compiler based on continuous-time Markov chains. Our method builds on the foundation of existing randomized Trotterization schemes, such as qDRIFT, while offering greater flexibility through continuous-time dependence. We derived and analyzed the system's governing equations, presented error bounds, and proved analytical bounds on the expected gate complexities under different models—both with perfect and imperfect gate applications.

Our results show that, under balanced schemes, the number of gates scales similarly to existing state-of-the-art methods like qDRIFT, but with added adaptability due to variable time-steps. This can potentially enhance performance in applications such as discretized adiabatic quantum computing. Importantly, our approach suggests that with proper parameter tuning, error rates can be kept minimal, ensuring reliable simulations over extended durations.

Further research could explore integrating our method with advanced strategies such as LCU (Linear Combination of Unitaries) or qSWIFT. Moreover, expanding the scope to randomized quantum optimization algorithms or discrete versions of adiabatic quantum computing might yield substantial practical improvements.
\newpage
\textbf{Acknowledgment:} This work was supported by the Belgian Fonds de la Recherche Scientifique - FNRS under Grants No. R.8015.21 (QOPT), O.0013.22 (EoS CHEQS), and B.D. is
a Aspirant grantee.

\newpage
\bibliographystyle{apsrev4-2} 
\bibliography{Article_Markov_Chain_Compiler}

\begin{thebibliography}{35}%
\makeatletter
\providecommand \@ifxundefined [1]{%
 \@ifx{#1\undefined}
}%
\providecommand \@ifnum [1]{%
 \ifnum #1\expandafter \@firstoftwo
 \else \expandafter \@secondoftwo
 \fi
}%
\providecommand \@ifx [1]{%
 \ifx #1\expandafter \@firstoftwo
 \else \expandafter \@secondoftwo
 \fi
}%
\providecommand \natexlab [1]{#1}%
\providecommand \enquote  [1]{``#1''}%
\providecommand \bibnamefont  [1]{#1}%
\providecommand \bibfnamefont [1]{#1}%
\providecommand \citenamefont [1]{#1}%
\providecommand \href@noop [0]{\@secondoftwo}%
\providecommand \href [0]{\begingroup \@sanitize@url \@href}%
\providecommand \@href[1]{\@@startlink{#1}\@@href}%
\providecommand \@@href[1]{\endgroup#1\@@endlink}%
\providecommand \@sanitize@url [0]{\catcode `\\12\catcode `\$12\catcode
  `\&12\catcode `\#12\catcode `\^12\catcode `\_12\catcode `\%12\relax}%
\providecommand \@@startlink[1]{}%
\providecommand \@@endlink[0]{}%
\providecommand \url  [0]{\begingroup\@sanitize@url \@url }%
\providecommand \@url [1]{\endgroup\@href {#1}{\urlprefix }}%
\providecommand \urlprefix  [0]{URL }%
\providecommand \Eprint [0]{\href }%
\providecommand \doibase [0]{https://doi.org/}%
\providecommand \selectlanguage [0]{\@gobble}%
\providecommand \bibinfo  [0]{\@secondoftwo}%
\providecommand \bibfield  [0]{\@secondoftwo}%
\providecommand \translation [1]{[#1]}%
\providecommand \BibitemOpen [0]{}%
\providecommand \bibitemStop [0]{}%
\providecommand \bibitemNoStop [0]{.\EOS\space}%
\providecommand \EOS [0]{\spacefactor3000\relax}%
\providecommand \BibitemShut  [1]{\csname bibitem#1\endcsname}%
\let\auto@bib@innerbib\@empty
\bibitem [{\citenamefont {Campbell}(2019)}]{campbellRandomCompilerFast2019}%
  \BibitemOpen
  \bibfield  {author} {\bibinfo {author} {\bibfnamefont {E.}~\bibnamefont
  {Campbell}},\ }\href {https://doi.org/10.1103/PhysRevLett.123.070503}
  {\bibfield  {journal} {\bibinfo  {journal} {Physical Review Letters}\
  }\textbf {\bibinfo {volume} {123}},\ \bibinfo {pages} {070503} (\bibinfo
  {year} {2019})},\ \bibinfo {note} {publisher: American Physical
  Society}\BibitemShut {NoStop}%
\bibitem [{\citenamefont
  {Feynman}(1982)}]{feynmanSimulatingPhysicsComputers1982}%
  \BibitemOpen
  \bibfield  {author} {\bibinfo {author} {\bibfnamefont {R.~P.}\ \bibnamefont
  {Feynman}},\ }\href {https://doi.org/10.1007/BF02650179} {\bibfield
  {journal} {\bibinfo  {journal} {International Journal of Theoretical
  Physics}\ }\textbf {\bibinfo {volume} {21}},\ \bibinfo {pages} {467}
  (\bibinfo {year} {1982})}\BibitemShut {NoStop}%
\bibitem [{\citenamefont {Lloyd}(1996)}]{lloydUniversalQuantumSimulators1996}%
  \BibitemOpen
  \bibfield  {author} {\bibinfo {author} {\bibfnamefont {S.}~\bibnamefont
  {Lloyd}},\ }\href {https://doi.org/10.1126/science.273.5278.1073} {\bibfield
  {journal} {\bibinfo  {journal} {Science}\ }\textbf {\bibinfo {volume}
  {273}},\ \bibinfo {pages} {1073} (\bibinfo {year} {1996})},\ \bibinfo {note}
  {publisher: American Association for the Advancement of Science}\BibitemShut
  {NoStop}%
\bibitem [{\citenamefont {Somma}\ \emph {et~al.}(2002)\citenamefont {Somma},
  \citenamefont {Ortiz}, \citenamefont {Gubernatis}, \citenamefont {Knill},\
  and\ \citenamefont {Laflamme}}]{sommaSimulatingPhysicalPhenomena2002}%
  \BibitemOpen
  \bibfield  {author} {\bibinfo {author} {\bibfnamefont {R.}~\bibnamefont
  {Somma}}, \bibinfo {author} {\bibfnamefont {G.}~\bibnamefont {Ortiz}},
  \bibinfo {author} {\bibfnamefont {J.~E.}\ \bibnamefont {Gubernatis}},
  \bibinfo {author} {\bibfnamefont {E.}~\bibnamefont {Knill}},\ and\ \bibinfo
  {author} {\bibfnamefont {R.}~\bibnamefont {Laflamme}},\ }\href
  {https://doi.org/10.1103/PhysRevA.65.042323} {\bibfield  {journal} {\bibinfo
  {journal} {Physical Review A}\ }\textbf {\bibinfo {volume} {65}},\ \bibinfo
  {pages} {042323} (\bibinfo {year} {2002})},\ \bibinfo {note} {publisher:
  American Physical Society}\BibitemShut {NoStop}%
\bibitem [{\citenamefont {Aspuru-Guzik}\ \emph {et~al.}(2005)\citenamefont
  {Aspuru-Guzik}, \citenamefont {Dutoi}, \citenamefont {Love},\ and\
  \citenamefont {Head-Gordon}}]{aspuru-guzikSimulatedQuantumComputation2005}%
  \BibitemOpen
  \bibfield  {author} {\bibinfo {author} {\bibfnamefont {A.}~\bibnamefont
  {Aspuru-Guzik}}, \bibinfo {author} {\bibfnamefont {A.~D.}\ \bibnamefont
  {Dutoi}}, \bibinfo {author} {\bibfnamefont {P.~J.}\ \bibnamefont {Love}},\
  and\ \bibinfo {author} {\bibfnamefont {M.}~\bibnamefont {Head-Gordon}},\
  }\href {https://doi.org/10.1126/science.1113479} {\bibfield  {journal}
  {\bibinfo  {journal} {Science}\ }\textbf {\bibinfo {volume} {309}},\ \bibinfo
  {pages} {1704} (\bibinfo {year} {2005})},\ \bibinfo {note} {publisher:
  American Association for the Advancement of Science}\BibitemShut {NoStop}%
\bibitem [{\citenamefont {Lucas}(2014)}]{lucasIsingFormulationsMany2014}%
  \BibitemOpen
  \bibfield  {author} {\bibinfo {author} {\bibfnamefont {A.}~\bibnamefont
  {Lucas}},\ }\bibfield  {journal} {\bibinfo  {journal} {Frontiers in Physics}\
  }\textbf {\bibinfo {volume} {2}},\ \href
  {https://doi.org/10.3389/fphy.2014.00005} {10.3389/fphy.2014.00005} (\bibinfo
  {year} {2014}),\ \bibinfo {note} {publisher: Frontiers}\BibitemShut {NoStop}%
\bibitem [{\citenamefont {Childs}\ \emph {et~al.}(2017)\citenamefont {Childs},
  \citenamefont {Kothari},\ and\ \citenamefont
  {Somma}}]{childsQuantumAlgorithmSystems2017}%
  \BibitemOpen
  \bibfield  {author} {\bibinfo {author} {\bibfnamefont {A.~M.}\ \bibnamefont
  {Childs}}, \bibinfo {author} {\bibfnamefont {R.}~\bibnamefont {Kothari}},\
  and\ \bibinfo {author} {\bibfnamefont {R.~D.}\ \bibnamefont {Somma}},\ }\href
  {https://doi.org/10.1137/16M1087072} {\bibfield  {journal} {\bibinfo
  {journal} {SIAM J. Comput.}\ }\textbf {\bibinfo {volume} {46}},\ \bibinfo
  {pages} {1920} (\bibinfo {year} {2017})}\BibitemShut {NoStop}%
\bibitem [{\citenamefont {Robert}\ \emph {et~al.}(2021)\citenamefont {Robert},
  \citenamefont {Barkoutsos}, \citenamefont {Woerner},\ and\ \citenamefont
  {Tavernelli}}]{robertResourceefficientQuantumAlgorithm2021}%
  \BibitemOpen
  \bibfield  {author} {\bibinfo {author} {\bibfnamefont {A.}~\bibnamefont
  {Robert}}, \bibinfo {author} {\bibfnamefont {P.~K.}\ \bibnamefont
  {Barkoutsos}}, \bibinfo {author} {\bibfnamefont {S.}~\bibnamefont
  {Woerner}},\ and\ \bibinfo {author} {\bibfnamefont {I.}~\bibnamefont
  {Tavernelli}},\ }\href {https://doi.org/10.1038/s41534-021-00368-4}
  {\bibfield  {journal} {\bibinfo  {journal} {npj Quantum Information}\
  }\textbf {\bibinfo {volume} {7}},\ \bibinfo {pages} {1} (\bibinfo {year}
  {2021})},\ \bibinfo {note} {publisher: Nature Publishing Group}\BibitemShut
  {NoStop}%
\bibitem [{\citenamefont {An}\ \emph {et~al.}(2023)\citenamefont {An},
  \citenamefont {Liu},\ and\ \citenamefont
  {Lin}}]{anLinearCombinationHamiltonian2023}%
  \BibitemOpen
  \bibfield  {author} {\bibinfo {author} {\bibfnamefont {D.}~\bibnamefont
  {An}}, \bibinfo {author} {\bibfnamefont {J.-P.}\ \bibnamefont {Liu}},\ and\
  \bibinfo {author} {\bibfnamefont {L.}~\bibnamefont {Lin}},\ }\href
  {https://doi.org/10.1103/PhysRevLett.131.150603} {\bibfield  {journal}
  {\bibinfo  {journal} {Physical Review Letters}\ }\textbf {\bibinfo {volume}
  {131}},\ \bibinfo {pages} {150603} (\bibinfo {year} {2023})},\ \bibinfo
  {note} {publisher: American Physical Society}\BibitemShut {NoStop}%
\bibitem [{\citenamefont {Zlokapa}\ and\ \citenamefont
  {Somma}(2024)}]{zlokapaHamiltonianSimulationLowenergy2024}%
  \BibitemOpen
  \bibfield  {author} {\bibinfo {author} {\bibfnamefont {A.}~\bibnamefont
  {Zlokapa}}\ and\ \bibinfo {author} {\bibfnamefont {R.~D.}\ \bibnamefont
  {Somma}},\ }\href {https://doi.org/10.22331/q-2024-08-27-1449} {\bibfield
  {journal} {\bibinfo  {journal} {Quantum}\ }\textbf {\bibinfo {volume} {8}},\
  \bibinfo {pages} {1449} (\bibinfo {year} {2024})},\ \bibinfo {note}
  {publisher: Verein zur Förderung des Open Access Publizierens in den
  Quantenwissenschaften}\BibitemShut {NoStop}%
\bibitem [{\citenamefont
  {Suzuki}(1990)}]{suzukiFractalDecompositionExponential1990}%
  \BibitemOpen
  \bibfield  {author} {\bibinfo {author} {\bibfnamefont {M.}~\bibnamefont
  {Suzuki}},\ }\href {https://doi.org/10.1016/0375-9601(90)90962-N} {\bibfield
  {journal} {\bibinfo  {journal} {Physics Letters A}\ }\textbf {\bibinfo
  {volume} {146}},\ \bibinfo {pages} {319} (\bibinfo {year}
  {1990})}\BibitemShut {NoStop}%
\bibitem [{\citenamefont {Suzuki}(1992)}]{suzukiFractalPathIntegrals1992}%
  \BibitemOpen
  \bibfield  {author} {\bibinfo {author} {\bibfnamefont {M.}~\bibnamefont
  {Suzuki}},\ }\href {https://doi.org/10.1016/0378-4371(92)90574-A} {\bibfield
  {journal} {\bibinfo  {journal} {Physica A: Statistical Mechanics and its
  Applications}\ }\textbf {\bibinfo {volume} {191}},\ \bibinfo {pages} {501}
  (\bibinfo {year} {1992})}\BibitemShut {NoStop}%
\bibitem [{\citenamefont {Berry}\ \emph {et~al.}(2007)\citenamefont {Berry},
  \citenamefont {Ahokas}, \citenamefont {Cleve},\ and\ \citenamefont
  {Sanders}}]{berryEfficientQuantumAlgorithms2007}%
  \BibitemOpen
  \bibfield  {author} {\bibinfo {author} {\bibfnamefont {D.~W.}\ \bibnamefont
  {Berry}}, \bibinfo {author} {\bibfnamefont {G.}~\bibnamefont {Ahokas}},
  \bibinfo {author} {\bibfnamefont {R.}~\bibnamefont {Cleve}},\ and\ \bibinfo
  {author} {\bibfnamefont {B.~C.}\ \bibnamefont {Sanders}},\ }\href
  {https://doi.org/10.1007/s00220-006-0150-x} {\bibfield  {journal} {\bibinfo
  {journal} {Communications in Mathematical Physics}\ }\textbf {\bibinfo
  {volume} {270}},\ \bibinfo {pages} {359} (\bibinfo {year}
  {2007})}\BibitemShut {NoStop}%
\bibitem [{\citenamefont {Poulin}\ \emph {et~al.}(2011)\citenamefont {Poulin},
  \citenamefont {Qarry}, \citenamefont {Somma},\ and\ \citenamefont
  {Verstraete}}]{poulinQuantumSimulationTimeDependent2011}%
  \BibitemOpen
  \bibfield  {author} {\bibinfo {author} {\bibfnamefont {D.}~\bibnamefont
  {Poulin}}, \bibinfo {author} {\bibfnamefont {A.}~\bibnamefont {Qarry}},
  \bibinfo {author} {\bibfnamefont {R.}~\bibnamefont {Somma}},\ and\ \bibinfo
  {author} {\bibfnamefont {F.}~\bibnamefont {Verstraete}},\ }\href
  {https://doi.org/10.1103/PhysRevLett.106.170501} {\bibfield  {journal}
  {\bibinfo  {journal} {Physical Review Letters}\ }\textbf {\bibinfo {volume}
  {106}},\ \bibinfo {pages} {170501} (\bibinfo {year} {2011})},\ \bibinfo
  {note} {publisher: American Physical Society}\BibitemShut {NoStop}%
\bibitem [{\citenamefont {Wallman}\ and\ \citenamefont
  {Emerson}(2016)}]{wallmanNoiseTailoringScalable2016}%
  \BibitemOpen
  \bibfield  {author} {\bibinfo {author} {\bibfnamefont {J.~J.}\ \bibnamefont
  {Wallman}}\ and\ \bibinfo {author} {\bibfnamefont {J.}~\bibnamefont
  {Emerson}},\ }\href {https://doi.org/10.1103/PhysRevA.94.052325} {\bibfield
  {journal} {\bibinfo  {journal} {Physical Review A}\ }\textbf {\bibinfo
  {volume} {94}},\ \bibinfo {pages} {052325} (\bibinfo {year} {2016})},\
  \bibinfo {note} {publisher: American Physical Society}\BibitemShut {NoStop}%
\bibitem [{\citenamefont {Knee}\ and\ \citenamefont
  {Munro}(2015)}]{kneeOptimalTrotterizationUniversal2015}%
  \BibitemOpen
  \bibfield  {author} {\bibinfo {author} {\bibfnamefont {G.~C.}\ \bibnamefont
  {Knee}}\ and\ \bibinfo {author} {\bibfnamefont {W.~J.}\ \bibnamefont
  {Munro}},\ }\href {https://doi.org/10.1103/PhysRevA.91.052327} {\bibfield
  {journal} {\bibinfo  {journal} {Physical Review A}\ }\textbf {\bibinfo
  {volume} {91}},\ \bibinfo {pages} {052327} (\bibinfo {year} {2015})},\
  \bibinfo {note} {publisher: American Physical Society}\BibitemShut {NoStop}%
\bibitem [{\citenamefont {Campbell}(2017)}]{campbellShorterGateSequences2017}%
  \BibitemOpen
  \bibfield  {author} {\bibinfo {author} {\bibfnamefont {E.}~\bibnamefont
  {Campbell}},\ }\href {https://doi.org/10.1103/PhysRevA.95.042306} {\bibfield
  {journal} {\bibinfo  {journal} {Physical Review A}\ }\textbf {\bibinfo
  {volume} {95}},\ \bibinfo {pages} {042306} (\bibinfo {year} {2017})},\
  \bibinfo {note} {publisher: American Physical Society}\BibitemShut {NoStop}%
\bibitem [{\citenamefont
  {Hastings}(2017)}]{hastingsWeightReductionQuantum2017}%
  \BibitemOpen
  \bibfield  {author} {\bibinfo {author} {\bibfnamefont {M.~B.}\ \bibnamefont
  {Hastings}},\ }\href@noop {} {\bibfield  {journal} {\bibinfo  {journal}
  {Quantum Info. Comput.}\ }\textbf {\bibinfo {volume} {17}},\ \bibinfo {pages}
  {1307} (\bibinfo {year} {2017})}\BibitemShut {NoStop}%
\bibitem [{\citenamefont {Childs}\ \emph {et~al.}(2019)\citenamefont {Childs},
  \citenamefont {Ostrander},\ and\ \citenamefont
  {Su}}]{childsFasterQuantumSimulation2019}%
  \BibitemOpen
  \bibfield  {author} {\bibinfo {author} {\bibfnamefont {A.~M.}\ \bibnamefont
  {Childs}}, \bibinfo {author} {\bibfnamefont {A.}~\bibnamefont {Ostrander}},\
  and\ \bibinfo {author} {\bibfnamefont {Y.}~\bibnamefont {Su}},\ }\href
  {https://doi.org/10.22331/q-2019-09-02-182} {\bibfield  {journal} {\bibinfo
  {journal} {Quantum}\ }\textbf {\bibinfo {volume} {3}},\ \bibinfo {pages}
  {182} (\bibinfo {year} {2019})},\ \bibinfo {note} {arXiv:1805.08385
  [quant-ph]}\BibitemShut {NoStop}%
\bibitem [{\citenamefont {Roland}\ and\ \citenamefont
  {Cerf}(2002)}]{rolandQuantumSearchLocal2002a}%
  \BibitemOpen
  \bibfield  {author} {\bibinfo {author} {\bibfnamefont {J.}~\bibnamefont
  {Roland}}\ and\ \bibinfo {author} {\bibfnamefont {N.~J.}\ \bibnamefont
  {Cerf}},\ }\href {https://doi.org/10.1103/PhysRevA.65.042308} {\bibfield
  {journal} {\bibinfo  {journal} {Physical Review A}\ }\textbf {\bibinfo
  {volume} {65}},\ \bibinfo {pages} {042308} (\bibinfo {year} {2002})},\
  \bibinfo {note} {publisher: American Physical Society}\BibitemShut {NoStop}%
\bibitem [{\citenamefont {Cunningham}\ and\ \citenamefont
  {Roland}(2024)}]{cunninghamEigenpathTraversalPoissonDistributed2024}%
  \BibitemOpen
  \bibfield  {author} {\bibinfo {author} {\bibfnamefont {J.}~\bibnamefont
  {Cunningham}}\ and\ \bibinfo {author} {\bibfnamefont {J.}~\bibnamefont
  {Roland}},\ }in\ \href {https://doi.org/10.4230/LIPIcs.TQC.2024.7} {\emph
  {\bibinfo {booktitle} {19th {Conference} on the {Theory} of {Quantum}
  {Computation}, {Communication} and {Cryptography} ({TQC} 2024)}}}\ (\bibinfo
  {publisher} {Schloss Dagstuhl – Leibniz-Zentrum für Informatik},\ \bibinfo
  {year} {2024})\ pp.\ \bibinfo {pages} {7:1--7:20}\BibitemShut {NoStop}%
\bibitem [{\citenamefont
  {Chakraborty}(2024)}]{chakrabortyImplementingAnyLinear2024}%
  \BibitemOpen
  \bibfield  {author} {\bibinfo {author} {\bibfnamefont {S.}~\bibnamefont
  {Chakraborty}},\ }\href {https://doi.org/10.22331/q-2024-10-10-1496}
  {\bibfield  {journal} {\bibinfo  {journal} {Quantum}\ }\textbf {\bibinfo
  {volume} {8}},\ \bibinfo {pages} {1496} (\bibinfo {year} {2024})},\ \bibinfo
  {note} {publisher: Verein zur Förderung des Open Access Publizierens in den
  Quantenwissenschaften}\BibitemShut {NoStop}%
\bibitem [{\citenamefont {Nakaji}\ \emph {et~al.}(2024)\citenamefont {Nakaji},
  \citenamefont {Bagherimehrab},\ and\ \citenamefont
  {Aspuru-Guzik}}]{nakajiHighOrderRandomizedCompiler2024}%
  \BibitemOpen
  \bibfield  {author} {\bibinfo {author} {\bibfnamefont {K.}~\bibnamefont
  {Nakaji}}, \bibinfo {author} {\bibfnamefont {M.}~\bibnamefont
  {Bagherimehrab}},\ and\ \bibinfo {author} {\bibfnamefont {A.}~\bibnamefont
  {Aspuru-Guzik}},\ }\href {https://doi.org/10.1103/PRXQuantum.5.020330}
  {\bibfield  {journal} {\bibinfo  {journal} {PRX Quantum}\ }\textbf {\bibinfo
  {volume} {5}},\ \bibinfo {pages} {020330} (\bibinfo {year} {2024})},\
  \bibinfo {note} {publisher: American Physical Society}\BibitemShut {NoStop}%
\bibitem [{\citenamefont {Berry}\ and\ \citenamefont
  {Childs}(2012)}]{berryBlackboxHamiltonianSimulation2012}%
  \BibitemOpen
  \bibfield  {author} {\bibinfo {author} {\bibfnamefont {D.~W.}\ \bibnamefont
  {Berry}}\ and\ \bibinfo {author} {\bibfnamefont {A.~M.}\ \bibnamefont
  {Childs}},\ }\href@noop {} {\bibfield  {journal} {\bibinfo  {journal}
  {Quantum Info. Comput.}\ }\textbf {\bibinfo {volume} {12}},\ \bibinfo {pages}
  {29} (\bibinfo {year} {2012})}\BibitemShut {NoStop}%
\bibitem [{\citenamefont {Berry}\ \emph
  {et~al.}(2015{\natexlab{a}})\citenamefont {Berry}, \citenamefont {Childs},\
  and\ \citenamefont {Kothari}}]{berryHamiltonianSimulationNearly2015}%
  \BibitemOpen
  \bibfield  {author} {\bibinfo {author} {\bibfnamefont {D.~W.}\ \bibnamefont
  {Berry}}, \bibinfo {author} {\bibfnamefont {A.~M.}\ \bibnamefont {Childs}},\
  and\ \bibinfo {author} {\bibfnamefont {R.}~\bibnamefont {Kothari}},\ }\href
  {https://doi.org/10.1109/FOCS.2015.54} {\bibfield  {journal} {\bibinfo
  {journal} {Proceedings - 2015 IEEE 56th Annual Symposium on Foundations of
  Computer Science, FOCS 2015}\ ,\ \bibinfo {pages} {792}} (\bibinfo {year}
  {2015}{\natexlab{a}})},\ \bibinfo {note} {place: Piscataway, NJ Publisher:
  Institute of Electrical and Electronics Engineers (IEEE)}\BibitemShut
  {NoStop}%
\bibitem [{\citenamefont {Berry}\ \emph
  {et~al.}(2015{\natexlab{b}})\citenamefont {Berry}, \citenamefont {Childs},
  \citenamefont {Cleve}, \citenamefont {Kothari},\ and\ \citenamefont
  {Somma}}]{berrySimulatingHamiltonianDynamics2015}%
  \BibitemOpen
  \bibfield  {author} {\bibinfo {author} {\bibfnamefont {D.~W.}\ \bibnamefont
  {Berry}}, \bibinfo {author} {\bibfnamefont {A.~M.}\ \bibnamefont {Childs}},
  \bibinfo {author} {\bibfnamefont {R.}~\bibnamefont {Cleve}}, \bibinfo
  {author} {\bibfnamefont {R.}~\bibnamefont {Kothari}},\ and\ \bibinfo {author}
  {\bibfnamefont {R.~D.}\ \bibnamefont {Somma}},\ }\href
  {https://doi.org/10.1103/PhysRevLett.114.090502} {\bibfield  {journal}
  {\bibinfo  {journal} {Physical Review Letters}\ }\textbf {\bibinfo {volume}
  {114}},\ \bibinfo {pages} {090502} (\bibinfo {year} {2015}{\natexlab{b}})},\
  \bibinfo {note} {publisher: American Physical Society}\BibitemShut {NoStop}%
\bibitem [{\citenamefont {Babbush}\ \emph {et~al.}(2018)\citenamefont
  {Babbush}, \citenamefont {Gidney}, \citenamefont {Berry}, \citenamefont
  {Wiebe}, \citenamefont {McClean}, \citenamefont {Paler}, \citenamefont
  {Fowler},\ and\ \citenamefont
  {Neven}}]{babbushEncodingElectronicSpectra2018}%
  \BibitemOpen
  \bibfield  {author} {\bibinfo {author} {\bibfnamefont {R.}~\bibnamefont
  {Babbush}}, \bibinfo {author} {\bibfnamefont {C.}~\bibnamefont {Gidney}},
  \bibinfo {author} {\bibfnamefont {D.~W.}\ \bibnamefont {Berry}}, \bibinfo
  {author} {\bibfnamefont {N.}~\bibnamefont {Wiebe}}, \bibinfo {author}
  {\bibfnamefont {J.}~\bibnamefont {McClean}}, \bibinfo {author} {\bibfnamefont
  {A.}~\bibnamefont {Paler}}, \bibinfo {author} {\bibfnamefont
  {A.}~\bibnamefont {Fowler}},\ and\ \bibinfo {author} {\bibfnamefont
  {H.}~\bibnamefont {Neven}},\ }\href
  {https://doi.org/10.1103/PhysRevX.8.041015} {\bibfield  {journal} {\bibinfo
  {journal} {Physical Review X}\ }\textbf {\bibinfo {volume} {8}},\ \bibinfo
  {pages} {041015} (\bibinfo {year} {2018})},\ \bibinfo {note} {publisher:
  American Physical Society}\BibitemShut {NoStop}%
\bibitem [{\citenamefont {Low}\ and\ \citenamefont
  {Chuang}(2019)}]{lowHamiltonianSimulationQubitization2019}%
  \BibitemOpen
  \bibfield  {author} {\bibinfo {author} {\bibfnamefont {G.~H.}\ \bibnamefont
  {Low}}\ and\ \bibinfo {author} {\bibfnamefont {I.~L.}\ \bibnamefont
  {Chuang}},\ }\href {https://doi.org/10.22331/q-2019-07-12-163} {\bibfield
  {journal} {\bibinfo  {journal} {Quantum}\ }\textbf {\bibinfo {volume} {3}},\
  \bibinfo {pages} {163} (\bibinfo {year} {2019})},\ \bibinfo {note}
  {publisher: Verein zur Förderung des Open Access Publizierens in den
  Quantenwissenschaften}\BibitemShut {NoStop}%
\bibitem [{\citenamefont {Babbush}\ \emph {et~al.}(2015)\citenamefont
  {Babbush}, \citenamefont {McClean}, \citenamefont {Wecker}, \citenamefont
  {Aspuru-Guzik},\ and\ \citenamefont
  {Wiebe}}]{babbushChemicalBasisTrotterSuzuki2015}%
  \BibitemOpen
  \bibfield  {author} {\bibinfo {author} {\bibfnamefont {R.}~\bibnamefont
  {Babbush}}, \bibinfo {author} {\bibfnamefont {J.}~\bibnamefont {McClean}},
  \bibinfo {author} {\bibfnamefont {D.}~\bibnamefont {Wecker}}, \bibinfo
  {author} {\bibfnamefont {A.}~\bibnamefont {Aspuru-Guzik}},\ and\ \bibinfo
  {author} {\bibfnamefont {N.}~\bibnamefont {Wiebe}},\ }\href
  {https://doi.org/10.1103/PhysRevA.91.022311} {\bibfield  {journal} {\bibinfo
  {journal} {Physical Review A}\ }\textbf {\bibinfo {volume} {91}},\ \bibinfo
  {pages} {022311} (\bibinfo {year} {2015})},\ \bibinfo {note} {publisher:
  American Physical Society}\BibitemShut {NoStop}%
\bibitem [{\citenamefont {Poulin}\ \emph {et~al.}(2015)\citenamefont {Poulin},
  \citenamefont {Hastings}, \citenamefont {Wecker}, \citenamefont {Wiebe},
  \citenamefont {Doberty},\ and\ \citenamefont
  {Troyer}}]{poulinTrotterStepSize2015}%
  \BibitemOpen
  \bibfield  {author} {\bibinfo {author} {\bibfnamefont {D.}~\bibnamefont
  {Poulin}}, \bibinfo {author} {\bibfnamefont {M.~B.}\ \bibnamefont
  {Hastings}}, \bibinfo {author} {\bibfnamefont {D.}~\bibnamefont {Wecker}},
  \bibinfo {author} {\bibfnamefont {N.}~\bibnamefont {Wiebe}}, \bibinfo
  {author} {\bibfnamefont {A.~C.}\ \bibnamefont {Doberty}},\ and\ \bibinfo
  {author} {\bibfnamefont {M.}~\bibnamefont {Troyer}},\ }\href@noop {}
  {\bibfield  {journal} {\bibinfo  {journal} {Quantum Info. Comput.}\ }\textbf
  {\bibinfo {volume} {15}},\ \bibinfo {pages} {361} (\bibinfo {year}
  {2015})}\BibitemShut {NoStop}%
\bibitem [{\citenamefont {Childs}\ \emph {et~al.}(2018)\citenamefont {Childs},
  \citenamefont {Maslov}, \citenamefont {Nam}, \citenamefont {Ross},\ and\
  \citenamefont {Su}}]{childsFirstQuantumSimulation2018}%
  \BibitemOpen
  \bibfield  {author} {\bibinfo {author} {\bibfnamefont {A.~M.}\ \bibnamefont
  {Childs}}, \bibinfo {author} {\bibfnamefont {D.}~\bibnamefont {Maslov}},
  \bibinfo {author} {\bibfnamefont {Y.}~\bibnamefont {Nam}}, \bibinfo {author}
  {\bibfnamefont {N.~J.}\ \bibnamefont {Ross}},\ and\ \bibinfo {author}
  {\bibfnamefont {Y.}~\bibnamefont {Su}},\ }\href
  {https://doi.org/10.1073/pnas.1801723115} {\bibfield  {journal} {\bibinfo
  {journal} {Proceedings of the National Academy of Sciences of the United
  States of America}\ }\textbf {\bibinfo {volume} {115}},\ \bibinfo {pages}
  {9456} (\bibinfo {year} {2018})}\BibitemShut {NoStop}%
\bibitem [{\citenamefont {Frigessi}\ and\ \citenamefont
  {Heidergott}(2011)}]{frigessiMarkovChains2011}%
  \BibitemOpen
  \bibfield  {author} {\bibinfo {author} {\bibfnamefont {A.}~\bibnamefont
  {Frigessi}}\ and\ \bibinfo {author} {\bibfnamefont {B.}~\bibnamefont
  {Heidergott}},\ }in\ \href {https://doi.org/10.1007/978-3-642-04898-2_347}
  {\emph {\bibinfo {booktitle} {International {Encyclopedia} of {Statistical}
  {Science}}}},\ \bibinfo {editor} {edited by\ \bibinfo {editor} {\bibfnamefont
  {M.}~\bibnamefont {Lovric}}}\ (\bibinfo  {publisher} {Springer},\ \bibinfo
  {address} {Berlin, Heidelberg},\ \bibinfo {year} {2011})\ pp.\ \bibinfo
  {pages} {772--775}\BibitemShut {NoStop}%
\bibitem [{\citenamefont {Yin}\ and\ \citenamefont
  {Zhang}(2013)}]{yinContinuousTimeMarkovChains2013}%
  \BibitemOpen
  \bibfield  {author} {\bibinfo {author} {\bibfnamefont {G.~G.}\ \bibnamefont
  {Yin}}\ and\ \bibinfo {author} {\bibfnamefont {Q.}~\bibnamefont {Zhang}},\
  }\href {https://doi.org/10.1007/978-1-4614-4346-9} {\emph {\bibinfo {title}
  {Continuous-{Time} {Markov} {Chains} and {Applications}}}},\ edited by\
  \bibinfo {editor} {\bibfnamefont {B.}~\bibnamefont {RozovskiĬ}}\ and\
  \bibinfo {editor} {\bibfnamefont {P.}~\bibnamefont {Glynn}},\ \bibinfo
  {series} {Stochastic {Modelling} and {Applied} {Probability}}, Vol.~\bibinfo
  {volume} {37}\ (\bibinfo  {publisher} {Springer},\ \bibinfo {address} {New
  York, NY},\ \bibinfo {year} {2013})\BibitemShut {NoStop}%
\bibitem [{\citenamefont {Fuchs}\ and\ \citenamefont {van~de
  Graaf}(1999)}]{fuchsCryptographicDistinguishabilityMeasures1999}%
  \BibitemOpen
  \bibfield  {author} {\bibinfo {author} {\bibfnamefont {C.}~\bibnamefont
  {Fuchs}}\ and\ \bibinfo {author} {\bibfnamefont {J.}~\bibnamefont {van~de
  Graaf}},\ }\href {https://doi.org/10.1109/18.761271} {\bibfield  {journal}
  {\bibinfo  {journal} {IEEE Transactions on Information Theory}\ }\textbf
  {\bibinfo {volume} {45}},\ \bibinfo {pages} {1216} (\bibinfo {year}
  {1999})},\ \bibinfo {note} {conference Name: IEEE Transactions on Information
  Theory}\BibitemShut {NoStop}%
\bibitem [{\citenamefont
  {Rastegin}(2012)}]{rasteginRelationsCertainSymmetric2012}%
  \BibitemOpen
  \bibfield  {author} {\bibinfo {author} {\bibfnamefont {A.~E.}\ \bibnamefont
  {Rastegin}},\ }\href {https://doi.org/10.1007/s10955-012-0569-8} {\bibfield
  {journal} {\bibinfo  {journal} {Journal of Statistical Physics}\ }\textbf
  {\bibinfo {volume} {148}},\ \bibinfo {pages} {1040} (\bibinfo {year}
  {2012})}\BibitemShut {NoStop}%
\end{thebibliography}%
\newpage
\onecolumngrid
\appendix
\section{Error for $Q$-node balanced scheme}\label{App:1}
If $\mcbs{R}(0)=\bs{0}$, then Equation~\eqref{Eq:Evol_ST_Sym} solves as
\begin{equation}
\begin{aligned}
    \mcbs{R}=-i&\int_0^te^{-\int_\tau^t\lambda d\tau'}\mcbs{T}\left(e^{-i\int_\tau^t\bs{\Pi}_{\vec{w}}\left(\left[\mcbs{H},\bs{\cdot}\right]\right)d\tau'}\right)\\&\left[\mcbs{H}-\bs{H}\otimes\bs{\mathds{1}},\bs{\rho}\otimes \bs{W}\right]d\tau.
\end{aligned}
\end{equation}
Let us bound each element. First the operator $\mcbs{T}\left(e^{-i\int_\tau^t\bs{\Pi}_{\vec{w}}\left(\left[\mcbs{H},\bs{\cdot}\right]\right)d\tau'}\right)$ is the operator that sends a density matrix $\mcbs{P}$ in a channel $\mcbs{C}(\tau\rightarrow \tau')$ such that at each time $\tau'$ the channel acts as
\begin{equation}
\partial_t\mcbs{P}=-i\bs{\Pi}_{\vec{w}}\left(\left[\mcbs{H},\mcbs{P}\right]\right).
\end{equation} 
We can look at the evolution of the norm of $\mcbs{P}$ at each time using the following identity (for sake of simplicity, we assume $\mcbs{P}$ positive definite, however a more cautious development allows us to extend the reasoning to any density matrix):
\begin{equation}
\norm{\mcbs{P}}_p^p=\tr\left(\mcbs{P}^p\right)\leftrightarrow \norm{\mcbs{P}}_p^{p-1}\partial_t\norm{\mcbs{P}}_p=\tr\left(\mcbs{P}^{p-1}\partial_t\mcbs{P}\right)
\end{equation}
Then:
\begin{equation}
\begin{aligned}
\norm{\mcbs{P}}_p^{p-1}\partial_t\norm{\mcbs{P}}_p&=\tr\left(\mcbs{P}^{p-1}\bs{\Pi}_{\vec{w}}\left(-i\left[\mcbs{H},\mcbs{P}\right]\right)\right)\\
&=\tr\left(-i\mcbs{P}^{p-1}\left[\mcbs{H},\mcbs{P}\right]\right)+ \tr\left(\mcbs{P}^{p-1}\left(\tr_Q\left(i\left[\mcbs{H},\mcbs{P}\right]\right)\otimes \bs{W}\right)\right)\\
&=\tr\left(\mcbs{P}^{p-1}\left(\tr_Q\left(\left[i\mcbs{H},\mcbs{P}\right]\right)\otimes \bs{W}\right)\right)\\
&\leq \norm{\mcbs{P}^{p-1}}_{\frac{p}{p-1}}\norm{\tr_Q\left(\left[i\mcbs{H},\mcbs{P}\right]\right)\otimes \bs{W}}_p\\
&\leq \norm{\mcbs{P}}^{p-1}_{p}    \norm{\tr_Q\left(\left[i\mcbs{H},\mcbs{P}\right]\right)}_p    \norm{ \bs{W}}_p\\
&\leq \norm{\mcbs{P}}^{p-1}_{p}    Q^{\frac{p-1}{p}}    \norm{\left[i\mcbs{H},\mcbs{P}\right]}_p   \norm{ \vec{w}}_p\\
&\leq \norm{\mcbs{P}}^{p-1}_{p}   2Q^{\frac{p-1}{p}}     \norm{\mcbs{H}}_\infty   \norm{\mcbs{P}}_p\norm{ \vec{w}}_p\\
\Rightarrow \partial_t\norm{\mcbs{P}}_p&\leq 2Q^{\frac{p-1}{p}}     \norm{\mcbs{H}}_\infty   \norm{ \vec{w}}_p\norm{\mcbs{P}}_p\\
\Rightarrow \norm{\mcbs{P}}_p(t)&\leq e^{\int_\tau^t 2Q^{\frac{p-1}{p}}     \norm{\mcbs{H}}_\infty   \norm{ \vec{w}}_p d\tau'}\norm{\mcbs{P}}_p(\tau)
\end{aligned}
\end{equation}
where we used~\cite{rasteginRelationsCertainSymmetric2012}. This directly gives
\begin{equation}
    \norm{\mcbs{T}\left(e^{-i\int_\tau^t\bs{\Pi}_{\vec{w}}\left(\left[\mcbs{H},\bs{\cdot}\right]\right)d\tau'}\right)}_p\leq e^{2Q^{\frac{p-1}{p}}\int_\tau^t\norm{\mcbs{H}}_\infty\norm{\vec{w}}_pd\tau'}.
\end{equation}
The following term can also be bounded as
\begin{equation}
\begin{aligned}
    &\norm{i\left[\mcbs{H}-\bs{H}\otimes\bs{\mathds{1}},\bs{\rho}\otimes \bs{W}\right]}_p\\\leq& 2\left(\sum_iw_i^p\norm{\bs{H}_i-\bs{H}}^p\right)^{\frac{1}{p}}\norm{\bs{\rho}}_p\\
    \leq&2\norm{\vec{w}}_{pq}\norm{\norm{\bs{H}_i-\bs{H}}_\infty}_{pr}\norm{\bs{\rho}}_p~\forall q,r\leq 1:\frac{1}{q}+\frac{1}{r}=1
\end{aligned}
\end{equation}
Note that in the last expression the first step shows that the source term contains an expression of the spread of the $\bs{H}_i$, that is proportional to the $p$-order centered moment of the distribution of Hamiltonians when $w_i=\frac{1}{Q}$. If $w_i=\frac{1}{Q}$ and $\norm{\bs{H}_i-\bs{H}}_\infty$ is bounded by $C~\forall i$, then this last term scales as $Q^{\frac{1}{p}-1}C$. For the following, we will choose $q=1$ and $r=\infty$.

Putting those blocks together
\begin{equation}
\begin{aligned}
    \norm{\mcbs{R}}_p\leq&\int_0^t e^{\int_\tau^t2Q^{\frac{p-1}{p}}\norm{\mcbs{H}}_\infty\norm{\vec{w}}_p-\lambda d\tau'}\\&2\norm{\vec{w}}_{p}\norm{\mcbs{H}-\bs{H}\otimes\bs{\mathds{1}}}_\infty\norm{\bs{\rho}}_pd\tau,
\end{aligned}
\end{equation}
Assuming that all terms are constant
\begin{equation}
    \norm{\mcbs{R}}_p\leq\frac{2\norm{\vec{w}}_{p}\norm{\mcbs{H}-\bs{H}\otimes\bs{\mathds{1}}}_\infty\norm{\bs{\rho}}_p}{\lambda-2Q^{\frac{p-1}{p}}\norm{\mcbs{H}}_\infty\norm{\vec{w}}_p},
\end{equation}
and finally
\begin{equation}
    \norm{\mcbs{E}_T-\mcbs{S}_T}_{p\rightarrow p}\leq \frac{4Q^{\frac{p-1}{p}}\norm{\vec{w}}_{p}\norm{\mcbs{H}-\bs{H}\otimes\bs{\mathds{1}}}_\infty\norm{\mcbs{H}}_\infty T}{\lambda-2Q^{\frac{p-1}{p}}\norm{\mcbs{H}}_\infty\norm{\vec{w}}_p}.
\end{equation}

\end{document}